# Construction of van der Waals magnetic tunnel junction using ferromagnetic layered dichalcogenide

Miho Arai[1], Rai Moriya*[,1], Naoto Yabuki[1], Satoru Masubuchi[1], Keiji Ueno[2], and Tomoki Machida*[,1,3]

[1] *Institute of Industrial Science, University of Tokyo, 4-6-1 Komaba, Meguro, Tokyo 153-8505, Japan*
[2] *Department of Chemistry, Graduate School of Science and Engineering, Saitama University, Saitama 338-8570, Japan*
[3] *Institute for Nano Quantum Information Electronics, University of Tokyo, 4-6-1 Komaba, Meguro, Tokyo 153-8505, Japan*

We investigate the micromechanical exfoliation and van der Waals (vdW) assembly of ferromagnetic layered dichalcogenide $Fe_{0.25}TaS_2$. The vdW interlayer coupling at the Fe-intercalated plane of $Fe_{0.25}TaS_2$ allows exfoliation of flakes. A vdW junction between the cleaved crystal surfaces is constructed by dry transfer method. We observe tunnel magnetoresistance in the resulting junction under an external magnetic field applied perpendicular to the plane, demonstrating spin-polarized tunneling between the ferromagnetic layered material through the vdW junction.

*E-mail: moriyar@iis.u-tokyo.ac.jp; tmachida@iis.u-tokyo.ac.jp

Recently, studies on two-dimensional (2D) materials such as graphene, hexagonal boron nitride (h-BN), and transition metal dichalcogenides (TMDs) have received considerable attention [1,2]. These materials exhibit layered structure in the bulk form, and the van der Waals (vdW) force connects each of the layers. Thus, these crystals can easily be exfoliated down to monolayer and used to construct heterostructures of different 2D materials connected by vdW force [3]. Metal, semiconductor, and insulator 2D materials have been studied, and various functional electronics and opto-electronics devices have been demonstrated with these materials [4-7]. Further, these 2D materials could be potentially applied in the field of spintronics [8]; for example, long-distance spin transport has been demonstrated in graphene/h-BN heterostructures [9], and spin polarized tunneling has been demonstrated through graphene [10] and h-BN [11,12]. In these experiments, the source for the spin-polarized electrons is a ferromagnetic metal fabricated by the evaporation technique; thus, the interface between the ferromagnetic and non-magnetic materials involves chemical bonding. On the other hand, the vdW interface does not require any chemical bonding at the junction, in principle. Therefore, layered ferromagnetic 2D materials could possibly be used for constructing vdW heterostructures with spintronic functions. Moreover, the vdW hetrostructure could provide another degree of freedom that has not been possible in the conventional spintronics device; such as controlling the interlayer twist [13,14] and building single-crystalline heterostructures free from the lattice mismatch problem [3]. These unique features of vdW heterostructure provide significant advances in spintronics applications. Several theoretical studies have explored ferromagnetic 2D materials [15,16]; however, vdW heterostructures have not been realized with ferromagnetic 2D materials.

In this study, we show the mechanical exfoliation and dry transfer of an Fe-intercalated TMD material, $Fe_{0.25}TaS_2$, and demonstrate the construction of a magnetic tunnel junction (MTJ) by connecting cleaved flakes with vdW force. The fabricated vdW-MTJ exhibited perpendicular magnetic anisotropy, and we demonstrated a tunnel magnetoresistance (TMR) signal of 6%. Our results present a significant step in the realization of spintronics devices with only vdW materials.

The crystal structure of $Fe_{0.25}TaS_2$ is schematically illustrated in Fig. 1(a). The Fe atoms are intercalated in the vdW gap of the $TaS_2$ crystals and maintain 2H-type layer stacking [17]. The material properties of $Fe_xTaS_2$ are well known in its bulk form, where $Fe_xTaS_2$ shows ferromagnetic ordering accompanied by strong perpendicular magnetic anisotropy when $x$ ranges between 0.15 and 0.45 [18,19]. The ferromagnetic ordering is strongest at $x = 0.25$ due to the formation of the Fe sub-lattice structure; thus, the highest ferromagnetic transition temperature $T_C$ of ~160 K can be achieved at this Fe composition. Previous studies revealed carrier-mediated Ruderman–Kittel–Kasuya–Yoshida (RKKY) interaction between Fe local moments [20] and the anomalous Hall effect of this material [21], suggesting spin polarization in the conduction electrons of $Fe_xTaS_2$. Further, the micromechanical exfoliation of $Fe_{0.28}TaS_2$ down to ~100 nm has been reported recently [22]. These results suggest that $Fe_xTaS_2$ could be suitable for constructing vdW heterostructures and for injecting spin-polarized electrons across the vdW interface.

To fabricate the vdW-MTJ, $Fe_{0.25}TaS_2$ crystals were synthesized using the iodine vapor transport method [23]. The fabricated $Fe_{0.25}TaS_2$ crystals are mechanically exfoliated to

~100-nm-thick flakes. Using the room temperature dry-transfer technique [24], cleaved surfaces of different flakes are subsequently connected with vdW force, as shown in Fig. 1(b). Note that the exfoliation of thinner flakes (on the order of tens of nanometers) is possible; however, these flakes are not large enough for fabricating the vdW junction. Thus, we used relatively thick flakes for the experiments presented in this study. Separate transport measurement on 70-nm-thick $Fe_{0.25}TaS_2$ flakes revealed the ferromagnetic transition temperature $T_C$ of ~160 K (Fig. S1 of supplementary material [25]), which is very close to the reported bulk property of this material [21]. Finally, for the transport measurement, Au (35 nm)/Ti (45 nm) electrodes are fabricated by electron-beam (EB) lithography and EB evaporation. All device fabrication processes were performed at room temperature without introducing any heat treatment; this is crucial to minimize oxidation of cleaved $Fe_{0.25}TaS_2$ surface and obtain good electrical contact to $Fe_{0.25}TaS_2$.

In Fig. 1(c), optical micrographs of the exfoliated flake and their vdW junction are shown. The area of vdW junction is 67.8 $\mu m^2$. The quality of the vdW interface is analyzed using cross-sectional transmission electron microscopy (TEM), as shown in Fig. 1(d). The TEM image revealed the layered structure of each $Fe_{0.25}TaS_2$ flake with a stacking period of 0.61 nm; this period is close to the reported value for the bulk material [18]. The vdW contact between the flakes is clearly visible in the TEM image [Fig. 1(d)]. The TEM reveals that vdW junction displays different TEM contrast from the flake; this behavior is attributed to the presence of a native oxide layer existing at the surface of cleaved $Fe_{0.25}TaS_2$. The presence of $Ta_2O_5$ oxide layer in the topmost layer of the exfoliated $Fe_{0.25}TaS_2$ surface is detected from TEM, energy dispersive X-ray (EDX) spectrometry, and X-ray photoelectron spectroscopy (XPS); these data are provided in Fig. S2 of supplementary material [25].

Here, we present the vertical transport properties of the $Fe_{0.25}TaS_2/Fe_{0.25}TaS_2$ vdW junction. The current-voltage (*I-V*) characteristic of the junction is evaluated by four-terminal measurements at 5 K, as shown in Fig. 2(a). The result exhibits a non-linear *I-V* curve; this non-linearity is also visible from the voltage dependence of the differential resistance *dV/dI* shown in the right axis of Fig. 2(a). Such non-linearity was not observed when the four-terminal measurement is performed on a single $Fe_{0.25}TaS_2$ flake. Moreover, the resistance (on the order of kilo-ohms) is much higher than the value expected from the resistance of the $Fe_{0.25}TaS_2$ flake. From experiments performed on other 100-nm-thick $Fe_{0.25}TaS_2$ flakes, we determined that the resistivity of $Fe_{0.25}TaS_2$ is 138 $\mu\Omega$-cm at 5 K. Therefore, we believe the observed non-linear *I-V* curve and the large resistance are due to vertical transport at the vdW junction. Further, the temperature dependence of junction resistance is also measured and compared with that for a 100-nm-thick $Fe_{0.25}TaS_2$ flake, as shown in Fig. 2(b). Both resistance values are normalized with the their values at 300 K. The junction resistance monotonically increases from room temperature to low temperature; such behavior is a reminiscent of the tunnel barrier. This temperature dependence is in contrast to that of $Fe_{0.25}TaS_2$ flake. The temperature dependence of the $Fe_{0.25}TaS_2$ flake displays metallic temperature dependence with a resistance drop below its ferromagnetic transition temperature (~160 K, in this case). Thus, the transport in the vdW junction is considerably different from that of the bulk, and tunneling transport of the thin insulating $Ta_2O_5$ barrier exists on the topmost layer of the each $Fe_{0.25}TaS_2$ surface.

The magnetic field dependence of the junction resistance $R_J$ is measured with the external magnetic field direction perpendicular to the plane, as shown in Fig. 3(a). The resistance–area ($R_JA$) products of the junction is determined as $R_JA$ = 580 kΩ-μm$^2$. We also present the anisotropic magnetoresistance (AMR) data measured by two-terminal measurement performed on the top and bottom Fe$_{0.25}$TaS$_2$ flakes [Fig. 3(b) and 3(c), respectively]. We observed magnetoresistance in the vdW junction. Moreover, because the top and bottom Fe$_{0.25}$TaS$_2$ flakes exhibit different coercive fields [Fig. 3(b, c)], the magnetoresistance signal around $B$ = ±5 T can be assigned to the switching between parallel and anti-parallel magnetization configurations at the vdW junction. The different coercive fields could originate from the coercive field distribution among different flakes with different thicknesses. We also note that the top Fe$_{0.25}$TaS$_2$ flake is highly strained because it only partially covers the ~100-nm-thick bottom Fe$_{0.25}$TaS$_2$ flake; such strain potentially modifies the coercive field of the flake. The observed signal cannot be explained by the AMR of the bulk region because the magnetoresistance jump of ~300 Ω in the junction resistance at $B$ = 5 T is much larger than the corresponding AMR signal of ~1 Ω. Therefore, the observed magnetoresistance signal is clear evidence of a TMR effect at the vdW junction. The existence of Ta$_2$O$_5$ tunnel barrier at the surface of Fe$_{0.25}$TaS$_2$ enable us spin-polarized tunneling thorough the vdW junction because Ta$_2$O$_5$ has been used as a tunnel barrier material in MTJs [26,27]; TMR has been also reported in these devices. Because the electrons of the intercalated Fe atoms are considered to be localized in Fe$_{0.25}$TaS$_2$, the tunneling occurs between the itinerant electrons in the Ta layer, which have finite polarization with the direction opposite to that of the Fe atoms [20,28].

In addition, the bias dependence of the magnetoresistance signal is measured, as shown in Fig. 4. The magnetoresistance signal decreases with increasing bias voltage, and the change is nearly symmetric for the polarity of the bias. Such bias dependence further suggests that the observed magnetoresistance signal is due to the TMR effect. To examine the reproducibility of the TMR effect, we fabricated three vdW junctions using the same fabrication method and revealed that all of these devices display a similar range of the TMR ratio (Fig. S3 of supplementary material [25]), thus demonstrating the reproducibility of the vdW-MTJ. On the other hand, the $R_JA$ of the junctions show variation between devices. Cross-sectional TEM of the junction revealed the existence of a non-contact region, in which $Fe_{0.25}TaS_2$ flakes are separated by hydrocarbons (Fig. S4 of supplementary material [25]). This non-contact region occurs due to the steps and terraces that frequently exist on the surface of the exfoliated flakes and does not contribute to the vertical transport in vdW junction Thus, although $R_JA$ varies, similar TMR has been observed among the fabricated vdW junctions.

In summary, we have demonstrated the fabrication of a vdW junction using flakes of ferromagnetic layered dichalcogenide, $Fe_{0.25}TaS_2$, and revealed that the vdW junction functions as a MTJ with perpendicular magnetic anisotropy. These results demonstrate the possibility of spin-polarized tunneling between layered materials connected with vdW interaction. Because only the dry-transfer technique is needed to create these junctions, our vdW-MTJ presents a number of opportunities for integrating various spintronic functions into complex vdW heterostructures. Furthermore, we expect that obtaining larger TMR ratio could be achievable in the future experiments by using recently developed vdW transfer method in inert atmosphere environment [29-31].

**Acknowledgement**

The authors are grateful to the Foundation for Promotion of Material Science and Technology of Japan (MST) for the TEM, EDX and XPS analysis. This work was partly supported by a Grant-in-Aid for Scientific Research on Innovative Areas, "Science of Atomic Layers" and "Nano Spin Conversion Science" from the Ministry of Education, Culture, Sports, Science and Technology (MEXT) and the Project for Developing Innovation Systems of MEXT and Grants-in-Aid for Scientific Research of the Japan Society for the Promotion of Science (JSPS).

**Figure captions**

Figure 1

(a) The crystal structure of $Fe_{0.25}TaS_2$ is schematically illustrated. The Fe atoms intercalate the $TaS_2$ substance with the formation of the 2×2 Fe sub-lattice structure and the $TaS_2$ planes are arranged as a 2H-type structure. (b) The fabrication of $Fe_{0.25}TaS_2/Fe_{0.25}TaS_2$ vdW junction. The cleaved flakes of $Fe_{0.25}TaS_2$ are connected to build junction by using the dry transfer method at room temperature. The Au/Ti electrodes are fabricated. (c) Optical micrograph of the cleaved flakes after mechanical exfoliation and their vdW junction. The thickness of both flakes is ~100 nm. (d) Cross-sectional TEM image of the vdW junction.

Figure 2

(a) Left axis: The four-terminal *I-V* curve measured in the vdW junction at 5 K. Right axis: The differential resistance *dV/dI*. (b) The temperature dependence of the four-terminal junction resistance measured at $I = 2$ μA is presented. For comparison, the temperature dependence of in-plane resistance measured on the single $Fe_{0.25}TaS_2$ flake with a thickness of 100 nm is also presented. Both curves are normalized to their values at 300 K.

Figure 3

(a) External magnetic field dependence of the four-terminal resistance measured from the vdW junction at 5 K. (b), (c) The two-terminal resistance measured on the top and bottom $Fe_{0.25}TaS_2$ flake, respectively. The contact configuration for the electrical measurement is illustrated on the right side of the figure. The measurement is performed in two-terminal resistance geometry. The current used for the measurement is $I = 2$ μA in all the cases.

Figure 4

Junction bias dependence of the TMR ratio obtained from the vdW junction in Fig. 3 measured at 5 K.

Figure 1

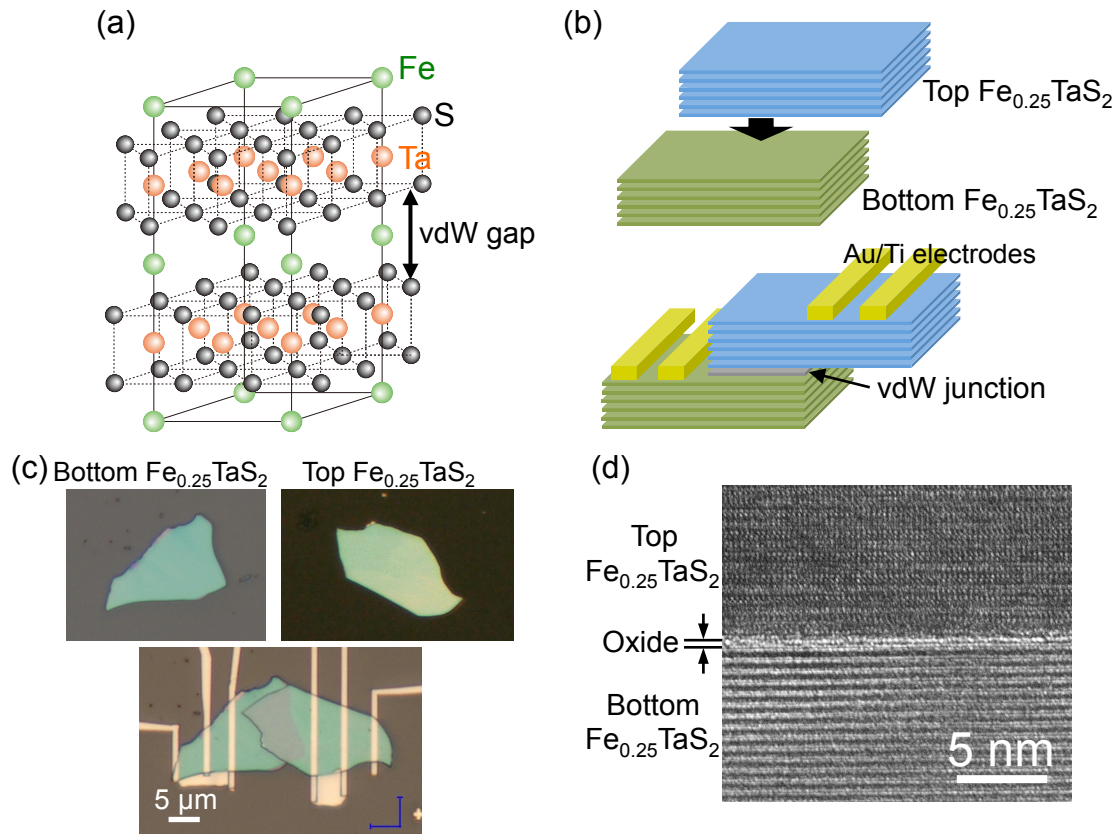

Figure 2

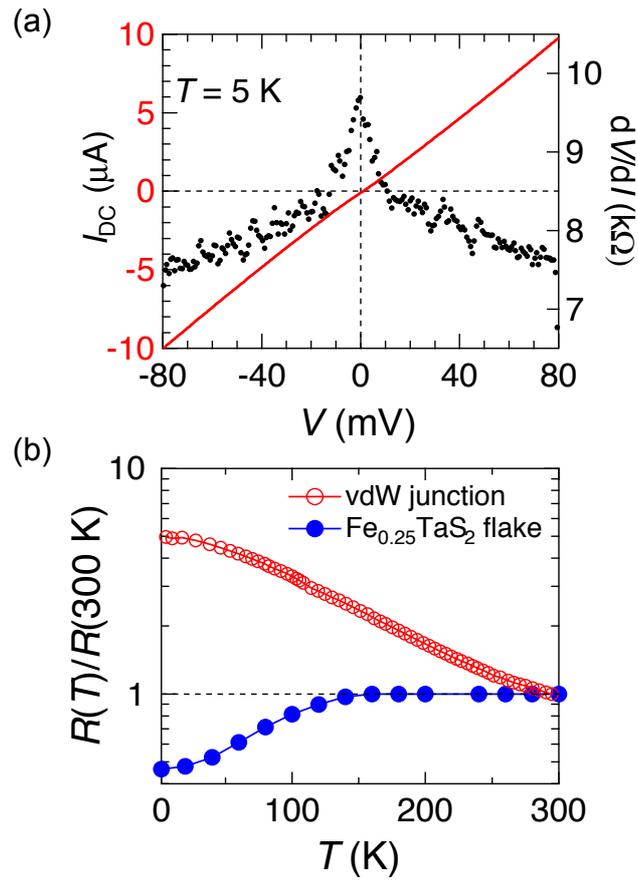

Figure 3

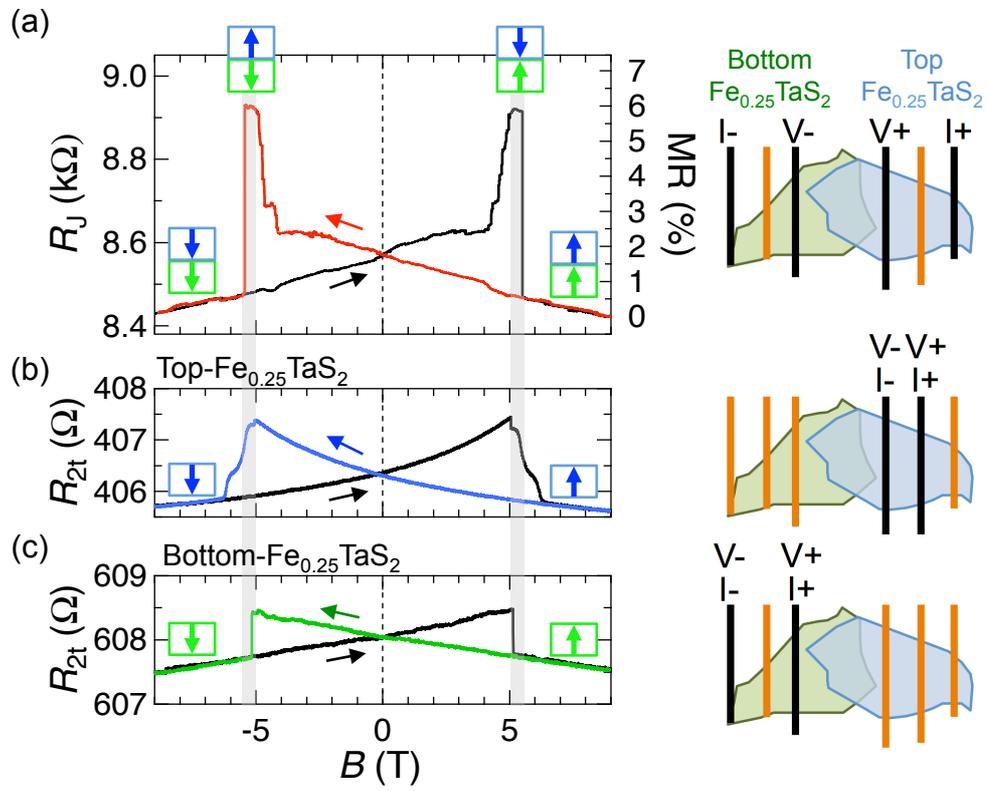

Figure 4

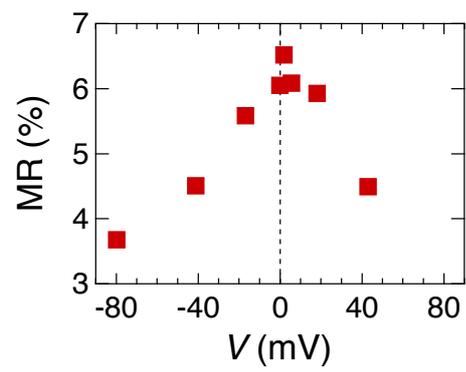